\pdfoutput=1
\documentclass[11pt]{article}

\usepackage[]{acl}

\usepackage{times}
\usepackage{latexsym}

\usepackage[T1]{fontenc}
\usepackage[utf8]{inputenc}
\usepackage{microtype}
\usepackage{inconsolata}

\usepackage{graphicx}
\usepackage{amsmath,amssymb,amsfonts}
\usepackage{booktabs}
\usepackage{float}
\usepackage{multirow}
\usepackage{url}
\usepackage{xurl}

\title{Revisiting Accent Classification with DistilHuBERT: Speaker-Disjoint Evaluation and Cross-Corpus Generalization}
  
\author{Anonymous Authors}

\author{
Xu He \\
Department of Electrical Engineering \\
Columbia University \\
New York, NY, USA \\
\texttt{xh2707@columbia.edu} \\
\And
Homayoon Beigi\\
Mechanical Engineering Dept. and\\
Electrical Engineering Dept.\\
Columbia University, NYC, NY, USA\\
and Recognition Technologies, Inc.\\
South Salem, NY, USA\\
 \texttt{homayoon.beigi@columbia.edu}
}

\begin{document}
\maketitle

\makeatletter
\def\@thefnmark{}\@footnotetext{Presented at the 9th International Conference on Natural Language and Speech Processing (ICNLSP 2026), University of Trento, Trento, Italy, September 25--26, 2026.}
\makeatother

\begin{abstract}
Recent accent classification systems can achieve high accuracy in standard in-corpus evaluations. However, when speakers overlap between the training and test sets, such results may overestimate model robustness compared with real-world deployment scenarios. This study revisits the robustness of DistilHuBERT-based accent classification by using L2-ARCTIC and multiple external English accent datasets. The comparison covers file-level and speaker-disjoint splits, along with balanced external-data training and teacher-student pseudo-labeling. The large performance gap between in-corpus and external evaluations suggests that file-level evaluation substantially overestimates performance, while speaker-disjoint evaluation can provide a more realistic estimate of the generalization to unseen speakers. In this study, models trained on external data alone do not reliably generalize to L2-ARCTIC. Among the tested configurations, balanced external-data training offers the most promising supervised setting. The pseudo-labeling experiments further indicate that confidence-based filtering alone is insufficient under domain mismatch. Overall, this study suggests that robust accent classification evaluation should simultaneously include speaker-disjoint evaluation and cross-corpus testing.
\end{abstract}


\section{Introduction}

\subsection{Background}

A speaker's native language or dialect can influence acoustic and prosodic patterns when speaking a second language such as English. Accent-related variation may appear in segmental and suprasegmental pronunciation features, such as vowel and consonant realization, stress, rhythm, and intonation \cite{b1}. The Hugging Face model \url{kaysrubio/accent-id-distilhubert-finetuned-l2-arctic2} is an open-source accent classifier based on DistilHuBERT \cite{b5}. This model was trained on the L2-ARCTIC corpus, which is a non-native English corpus containing recordings of speakers with six first-language backgrounds: Arabic, Hindi, Korean, Mandarin, Spanish, and Vietnamese \cite{b3,b8}.

However, the preliminary evaluation shows that the robustness of this model outside the original dataset is limited. Specifically, the model achieves near-perfect accuracy on L2-ARCTIC under an in-corpus file-level setting, but performs substantially worse on external datasets. One possible explanation is that the model may partially rely on the speaker-specific acoustic characteristics within the L2-ARCTIC dataset, rather than the common characteristics of a certain type of accent \cite{b2}.

To address this issue, this study revisits DistilHuBERT-based accent classification under speaker-disjoint and cross-corpus evaluation settings, and analyzes whether balanced external-data training and pseudo-labeling improve robustness beyond standard file-level evaluation \cite{b2,b8}.

\subsection{Motivation and Contributions}

The evaluation of the public Hugging Face baseline provides an example of a broader issue: high file-level accuracy alone is not sufficient evidence of robust accent classification. When the same speaker's voice samples are present in both the training and test sets, the model may utilize speaker-specific acoustic patterns rather than accent-level representations. This issue motivates a more rigorous evaluation of accent classification based on DistilHuBERT under speaker-disjoint and cross-corpus settings \citep{b2}.

This study makes four main contributions: (1) it reproduces and externally evaluates a public DistilHuBERT-based L2-ARCTIC accent-classification baseline \citep{b3,b8}, indicating that high in-corpus accuracy does not necessarily imply robust cross-corpus performance; (2) it constructs a speaker-disjoint evaluation protocol for L2-ARCTIC accent classification and compares it with file-level evaluation to examine the effect of speaker overlap on reported accuracy; (3) it introduces multiple external English accent datasets and evaluates an expanded eight-class setting, covering Arabic, Hindi, Korean, Mandarin, Spanish, Vietnamese, British Isles English, and Singaporean English; and (4) it evaluates balanced external-data construction and teacher-student pseudo-labeling to analyze whether external speech data can improve robustness under domain mismatch \citep{b9}.


\section{Related Work}

\subsection{Accent Classification and Non-Native English Speech}

Due to its close relationship with speech recognition, pronunciation assessment, and accent-aware spoken language technologies, automatic accent classification is receiving increasing attention.  Previous studies have shown that accent-related information can be obtained from the acoustic, phonetic,  and prosodic characteristics of speech \cite{b19,b20,b22}.

L2-ARCTIC is one of the commonly used corpora for non-native English speech research \cite{b3}. This dataset contains recordings from English speakers with six different first language backgrounds: Arabic, Hindi, Korean, Mandarin, Spanish, and Vietnamese. Moreover, in this dataset, speaker identities and accent labels are both available, so L2-ARCTIC is suitable for controlled accent-related experiments.

Other accent-related corpora have also been proposed. AccentDB, for example, was introduced as a database of non-native English accents for accent classification and accent-aware speech processing \cite{b19}. More recent studies have further explored accent classification using neural and pre-trained speech models \cite{b20}. These studies show that accent categories can be predicted from speech signals, but many of them focus primarily on classification accuracy rather than on whether the reported accuracy remains stable under speaker-disjoint or cross-corpus evaluation.

The publicly available Hugging Face baseline follows the research approach of fine-tuning DistilHuBERT on L2-ARCTIC for six-class accent classification \cite{b8}. Although such publicly available reports have demonstrated excellent performance within the corpus, their robustness under speaker-disjoint and external evaluation settings remains insufficiently studied.

\subsection{Speaker-Disjoint Evaluation and Cross-Corpus Generalization}

In accent classification research, a major challenge is to ensure that the reported performance can reflect the model's ability to learn accent-level patterns, rather than merely memorizing the specific characteristics of the speaker. The issue of speaker-disjoint evaluation is precisely related to this challenge. When the same speaker's utterances appear simultaneously in the training set and the test set, a model may utilize speaker identity cues instead of learning accent representations.


This issue has been discussed  in the context of non-native English automatic speech recognition. \citet{b2} distinguished between speaker-dependent and speaker-disjoint evaluation settings and showed the importance of evaluating generalization to unseen speakers. Although their work focused on ASR rather than accent classification, it highlights a broader concern about speaker memorization and evaluation validity.

Cross-corpus evaluation is an even more challenging setting. Differences in recording conditions, corpus design, speaker demographics, speaking style, and label definitions may lead to substantial domain mismatch between different datasets.  Models that perform well within a single corpus exhibit a significant performance decline when evaluated on external speech sources. Related research on robust automatic speech recognition (ASR) for accents also indicates that unseen accents and domain mismatches remain important factors contributing to performance degradation \cite{b21, b22}.

Prior work has investigated accent classification, speaker-disjoint evaluation, and cross-corpus robustness in speech processing. However, their combined impact has received less attention in DistilHuBERT-based accent classification. This study therefore jointly evaluates file-level, speaker-disjoint, and cross-corpus settings, and analyzes how speaker overlap and corpus dependence may inflate reported performance.

\subsection{Self-Supervised and Semi-Supervised Learning for Accent-Related Tasks}

Self-supervised speech representation learning has substantially improved the performance of downstream speech processing tasks. HuBERT learns context-dependent speech representations by predicting masked hidden units \cite{b4}, while wav2vec 2.0 indicates that large-scale self-supervised pre-training can reduce the need for labeled data in speech recognition and classification tasks \cite{b6}. DistilHuBERT retains most of the representation capabilities of HuBERT while further reducing computational costs through layer-wise distillation \cite{b5}.

Self-supervised representations have also been used for accent recognition and accent-robust speech recognition (ASR). \citet{b23} used self-supervised pre-training methods for both accent recognition and accented speech recognition. \citet{b21} studied accent-robust speech recognition using supervised and unsupervised wav2vec embeddings. These studies suggest that self-supervised speech models can encode accent-relevant information, motivating the use of DistilHuBERT as the backbone model in this study.

Semi-supervised learning provides another possible direction in cases where reliable accent annotations are limited. Pseudo-labeling is a simple and widely used approach in which a teacher model assigns labels to unlabeled samples, and high-confidence predictions are used to train a student model \cite{b9}. In accent classification, this method is attractive because external speech data is relatively easy to collect, while reliable accent annotations are often incomplete or contradictory.

Whereas much prior work focuses on improving accent-classification accuracy or accent-aware automatic speech recognition (ASR), this work focuses on evaluating robustness. Rather than introducing a new neural architecture, this study examines whether commonly reported accent-classification performance remains stable under stricter evaluation conditions. Specifically, this study explores evaluation methods at the file-level, speaker-disjoint, and cross-corpus settings within a unified DistilHuBERT framework, while also examining balanced external data training and the pseudo-labeling method.


\section{Methodology}

\subsection{Experimental Framework}

\begin{figure}[t] 
\centering 
\includegraphics[width=0.95\linewidth]{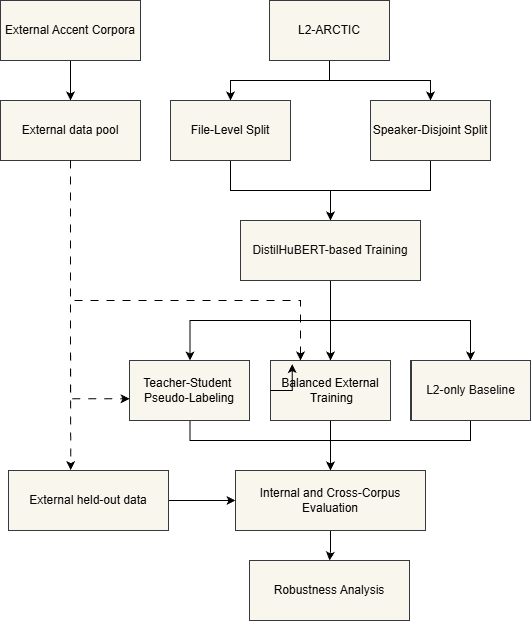} \caption{Overview of the speaker-disjoint and cross-corpus evaluation framework. Solid arrows indicate supervised training and evaluation, while dashed arrows indicate the use of external data for balanced training, pseudo-labeling, and held-out external evaluation.} 
\label{fig:framework} 
\end{figure}

Fig.~\ref{fig:framework} summarizes the experimental framework. The study first compares file-level and speaker-disjoint evaluation on L2-ARCTIC. Subsequently, external accent corpora are incorporated through balanced external data training, teacher-student pseudo-labeling, and held-out cross-corpus evaluation.

\subsection{Task Definition}

The task is defined as English accent classification. Given an input waveform $x_i$, the classifier predicts an accent label $y_i$ from a predefined label set. The original six-class setup uses the first language background represented in L2-ARCTIC:

\begin{equation} \begin{split} \mathcal{Y}_{6}=\{& \text{Arabic}, \text{Hindi}, \text{Korean},\\ &\text{Mandarin}, \text{Spanish}, \text{Vietnamese} \}. \end{split} \end{equation}

In the expanded external settings, two additional categories of English accent have been added:

\begin{equation} \begin{split} \mathcal{Y}_{8}=\mathcal{Y}_{6}\cup\{& \text{British Isles English},\\ &\text{Singaporean English} \}. \end{split} \end{equation}

All classifiers use DistilHuBERT as the acoustic encoder \citep{b5}, followed by a classification head trained with cross-entropy loss.

\subsection{Dataset and External Data Construction} 

The data used in this study consist of the in-domain L2-ARCTIC corpus and an external speech pool constructed from multiple public accent-related datasets. The in-domain corpus is L2-ARCTIC \citep{b3}, containing recordings from 24 non-native English speakers, covering six first language backgrounds, with four speakers in each category. The post-processed metadata contains 26,867 utterances, with approximately 4,300 to 4,500 utterances per category.

The external speech pool is used for baseline external evaluation, balanced expanded-label training, and held-out cross-corpus evaluation. The external sources include Common Accent, Common Native, CommonVoice Accent Test, English Dialects, MNSC-v1, MNSC-v1-extend, ESLTTS, GLOBE, and GLOBE\_V2 \citep{b10,b11,b12,b13,b14,b15,b16,b17,b18}. All audio files are converted to mono-channel 16 kHz waveform format to match the input format of DistilHuBERT.

When the label name and the dataset description clearly match one of the target accent categories, the external labels will be mapped to the target label space.  Samples with ambiguous, unsupported, or overly broad labels were excluded from the comparable external evaluation set. The second-level external evaluation set was held out from supervised training, development, testing, balanced external sampling, and pseudo-label generation to ensure an independent held-out evaluation.

The raw external data pool is highly imbalanced across labels and source datasets, so it should not be directly used as the final training set. Instead, a balanced external subset is constructed for the expanded eight-class setting. British Isles English and Singaporean English are selected as the two additional categories because they have sufficient samples for controlled balanced training. In Round 2, each new category contributes 3,000 training samples, 375 development samples, and 375 test samples. For fold 0, the final balanced training set contains 24,186 speech instances, with each of the six L2-ARCTIC categories contributing approximately 2,900 to 3,100 training samples.

\subsection{Speaker-Disjoint Evaluation}

Random file-level split may cause utterances from the same speaker to be included in both the training set and the test set simultaneously. In this case, the high accuracy rate may partly reflect speaker-specific cues or corpus artifacts, rather than the accent-level generalization. This issue is particularly important for L2-ARCTIC, as each accent category only contains four speakers.

\begin{table}[t]
\centering
\scriptsize
\setlength{\tabcolsep}{2pt}
\begin{tabular*}{\columnwidth}{@{\extracolsep{\fill}}lccc@{}}
\toprule
Setting & \shortstack{Speaker\\overlap} & \shortstack{External\\use} & Classes \\
\midrule
File-level baseline & Yes & No & 6 \\
Speaker-disjoint baseline & No & No & 6 \\
Balanced external training & No & Yes & 8 \\
Semi-supervised training & No & Pseudo-labels & 6/8 \\
Second-level external test & N/A & Held out & 8 \\
\bottomrule
\end{tabular*}
\caption{Evaluation and training protocols used in this study.}
\label{tab:split_protocols}
\end{table}

To achieve a more rigorous evaluation setup, this study constructed a four-fold speaker-disjoint protocol. In each round of testing, one speaker from each of the six accent categories was reserved for testing, while the remaining three speakers were used for training.  Therefore, each round of testing included six speakers, and each speaker was not present in both the training and testing sets simultaneously.

Table~\ref{tab:split_protocols} summarizes the main training and evaluation protocols used in this study.

\subsection{Teacher-Student Pseudo-Labeling}

Semi-supervised learning is implemented through teacher-student pseudo-labeling \cite{b9}. Two types of teacher models are employed: the first is a six-class DistilHuBERT model trained solely on speaker-disjoint L2-ARCTIC data; the second is an eight-class model trained on L2-ARCTIC plus balanced British Isles English and Singaporean English data.

For an external sample $x_j$, the teacher predicts a probability distribution $p_T(y|x_j)$, and the pseudo-labels are assigned in the following manner: 

\begin{equation}
\hat{y}_j=\arg\max_{y \in \mathcal{Y}} p_T(y|x_j).
\end{equation}

Only when the maximum confidence level of the sample exceeds the threshold $\tau$ will the sample be retained:

\begin{equation}
\max_{y \in \mathcal{Y}} p_T(y|x_j) \geq \tau.
\end{equation}

In the main semi-supervised setup, $\tau$ is set to 0.95, and each predicted category is limited to retain up to 500 pseudo-label samples.  This per-class category limit is intended to reduce the risk of class imbalance in the pseudo-label set.

The student model is trained on the union of the supervised training set and the selected pseudo-label samples:

\begin{equation}
\mathcal{D}_{student}=\mathcal{D}_{sup}\cup\mathcal{D}_{pseudo}.
\end{equation}

Here, $\mathcal{D}_{sup}$ denotes the supervised training set, $\mathcal{D}_{pseudo}$ denotes the selected external samples with pseudo labels, and $\mathcal{D}_{student}$ denotes the final training set used for the student model.

The second-level external evaluation set is excluded from supervised training and pseudo-label generation to  ensure that it serves as an independent benchmark dataset.

\section{Experiments and Results}

\subsection{Experimental Setup}

All experiments used DistilHuBERT as the acoustic encoder \cite{b5}. The publicly available Hugging Face baseline was used as a reference baseline \cite{b8}. For the newly trained models, the base checkpoint was \texttt{ntu-spml/distilhubert}, and the classification head was reinitialized according to the target label space.

The experiment aims to answer four main questions: (1) whether file-level evaluation overestimates accent-classification performance when speakers overlap between training and test sets; (2) how performance changes under speaker-disjoint evaluation; (3) whether balanced external-data training improves cross-corpus robustness; and (4) whether confidence-based pseudo-labeling provides additional robustness in the case of domain mismatch.

The experimental process consists of three stages. Round 1 is the diagnostic stage, comparing file-level training, speaker-disjoint training, external-data training, and external-only training. Round 2  involves retraining the most relevant supervised settings through balanced external-data construction and with the same training configuration as the public Hugging Face baseline. The final stage evaluates teacher-student pseudo-labeling using high-confidence predictions from trained teacher models. Throughout the entire pipeline, in addition to the public Hugging Face reference model, this study also trained 31 new DistilHuBERT models.

For Round 2 and semi-supervised student models, the training configuration is set as close as possible to the public baseline. The learning rate is $5\times10^{-5}$, the scheduler is linear, the warmup ratio is 0.1, and the number of epochs is 10. Due to GPU memory constraints, however, batch size for each device is set to 4, the gradient accumulation step is 2, and thus the actual effective batch size was 8.  The audio input is truncated or padded to the maximum of 8 seconds, and the feature encoder is frozen during training.

Four accuracy measures are reported. L2 accuracy denotes evaluation on the held-out L2-ARCTIC test split. External comparable accuracy denotes evaluation on external samples whose labels are mapped to the corresponding target label space. Combined accuracy is computed over the union of the L2-ARCTIC test set and the comparable external test samples. Level-2 external accuracy denotes evaluation on the held-out external set that is excluded from supervised training, balanced external sampling, and pseudo-label generation.

\subsection{Baseline Reproduction Check}

The first experiment verifies whether high in-corpus performance can be reproduced under a file-level L2-ARCTIC setting. As shown in Table~\ref{tab:baseline_reproduction}, the replicated file-level baseline achieves 99.76\% internal L2 accuracy, but the external accuracy is only 44.52\%. This result indicates that the strong in-corpus result is reproducible, but it does not alone demonstrate its robust cross-corpus generalization ability.

\begin{table}[t]
\centering
\small
\setlength{\tabcolsep}{4pt}
\begin{tabular}{lcc}
\toprule
Model & Internal L2 Acc. (\%) & External Acc. (\%) \\
\midrule
Baseline model & 99.76 & 44.52 \\
\bottomrule
\end{tabular}
\caption{Internal and external evaluation of the reproduced file-level baseline model.}
\label{tab:baseline_reproduction}
\end{table}

This gap motivates this study to adopt more rigorous speaker-disjoint evaluation and external held-out testing. Therefore, in subsequent experiments, file-level performance is regarded as an upper-bound diagnostic setting rather than as sufficient evidence of real-world robustness.

\subsection{Round 1 Diagnostic Experiments}

Round 1 compares file-level evaluation, speaker-disjoint evaluation, external-data training, and external-only training. Table~\ref{tab:first_round_summary} summarizes the main results.

\begin{table*}[t]
\centering
\small
\setlength{\tabcolsep}{4pt}
\begin{tabular}{llccc}
\toprule
Model & Labels & Training data & L2 Acc. (\%) & Ext. Acc. (\%) \\
\midrule
Hugging Face baseline & 6 & L2-ARCTIC & 91.65 & 58.96 \\
File-level baseline & 6 & L2-ARCTIC & 100.00 & 53.76 \\
Speaker-disjoint baseline & 6 & L2-ARCTIC & $81.40 \pm 3.62$ & $61.27 \pm 15.49$ \\
L2 + external & 6 & L2-ARCTIC + external & $81.09 \pm 4.10$ & $100.00 \pm 0.00$ \\
External only & 6 & external only & $16.35 \pm 0.00$ & $100.00 \pm 0.00$ \\
L2 + external & expanded & L2-ARCTIC + external & $83.01 \pm 2.80$ & $98.55 \pm 1.00$ \\
External only & expanded & external only & $14.28 \pm 0.03$ & $94.22 \pm 0.82$ \\
\bottomrule
\end{tabular}
\caption{Round 1 evaluation results on official L2-ARCTIC and external comparable samples. Fold-based settings are reported as mean $\pm$ standard deviation across four folds.}
\label{tab:first_round_summary}
\end{table*}

The file-level baseline model achieved an accuracy rate of 100.00\% on the official L2-ARCTIC test split, while the speaker-disjoint L2-only model reaches $81.40 \pm 3.62$\%. This difference supports the core hypothesis: when speaker overlap is present, file-level evaluation may lead to an overestimation of reported performance.

When evaluated back on L2-ARCTIC, the setting that only used external data performed poorly, with its official L2 accuracy being lower than 17\%. This indicates that external data cannot simply replace the supervision provided by in-domain L2-ARCTIC. At the same time, the high external comparable scores obtained by external-data settings should be interpreted cautiously, as the original external data set is severely imbalanced and may contain source-specific patterns. These results motivate the balanced external construction used in Round 2.

Table~\ref{tab:fold_level_exp2_exp5} compares the two most relevant supervised Round 1 settings at the fold level. The expanded L2+external setting improves over the L2-only speaker-disjoint baseline on all four folds. Across the four matched folds, the expanded setting improves L2 accuracy by an average of 1.61 percentage points (95\% CI: 0.16--3.07). Given the small number of folds, this interval should be interpreted cautiously. Therefore, Round 2 focuses on two supervised settings: the speaker-disjoint baseline using only L2, and a balanced expanded-label model constructed using L2-ARCTIC and controlled external data.

\begin{table}[!t]
\centering
\small
\setlength{\tabcolsep}{4pt}
\begin{tabular}{lcc}
\toprule
Fold & L2-only (\%) & Expanded L2+external (\%) \\
\midrule
0 & 86.80 & 87.10 \\
1 & 80.15 & 82.50 \\
2 & 79.25 & 80.95 \\
3 & 79.40 & 81.50 \\
\midrule
Mean & 81.40 & 83.01 \\
\bottomrule
\end{tabular}
\caption{Fold-level official L2 accuracy for the two most relevant Round 1 supervised settings.}
\label{tab:fold_level_exp2_exp5}
\end{table}

\subsection{Round 2 Balanced External Training}

Round 2 evaluates whether the most promising supervised direction from Round 1 remains useful after explicit external-data balancing. The expanded model added two new categories, namely British Isles English and Singaporean English, as both have sufficient external speech samples for controlled sampling. Each new category contributes 3,000 training samples, 375 development samples, and 375 test samples. The aim was to reduce the impact of label imbalance and prevent the classifier from being dominated by the largest external data source.

\begin{table*}[t]
\centering
\small
\setlength{\tabcolsep}{4pt}
\begin{tabular}{lccccc}
\toprule
Model & Fold & L2 Acc. (\%) & Ext. Acc. (\%) & Comb. Acc. (\%) & L2-Ext. Acc. (\%) \\
\midrule
Round 2 L2-only & 0 & 86.20 & 71.05 & 85.13 & 42.00 \\
Round 2 L2-only & 1 & 74.60 & 47.37 & 72.68 & -- \\
Round 2 L2-only & 2 & 80.10 & 63.16 & 78.90 & -- \\
Round 2 L2-only & 3 & 80.30 & 67.11 & 79.37 & -- \\
Round 2 L2 + British/Singaporean & 0 & 87.90 & -- & 81.97 & 68.00 \\
Round 2 L2 + British/Singaporean & 1 & 76.20 & -- & 70.82 & -- \\
\bottomrule
\end{tabular}
\caption{Round 2 evaluation results on L2-ARCTIC, external comparable samples, and the second-level external evaluation set. Dashes indicate settings where the external comparable accuracy is not directly comparable because the label space or test composition differs.}
\label{tab:round2_results}
\end{table*}

Table~\ref{tab:round2_results} summarizes the completed Round 2 models. The L2-only speaker-disjoint models provide a supervised baseline across four folds. The expanded L2+British/Singaporean accent setting completed two rounds. Due to limited data availability after label mapping, filtering, and class-balancing constraints were applied to the external corpora, the expanded Round 2 setting could only be constructed reliably for folds 0 and 1. As shown in Table~\ref{tab:round2_results}, on fold 0, the expanded model improves L2 test accuracy from 86.20\% to 87.90\% and improves the second-level external accuracy from 42.00\% to 68.00\%. On fold 1, the expanded model also improves L2 accuracy from 74.60\% to 76.20\%, although the combined accuracy remains lower than the fold-0 setting. These results indicate that, in some cases, balanced external data training can enhance the model's robustness, but its improvement effect is not uniform enough to claim a decisive improvement in all conditions.

The second-level external evaluation set is particularly important because it is not included in the supervised training, balanced external sampling, and pseudo-label generation processes. Therefore, it can more rigorously test whether the model has learned transferable accent-related knowledge rather than source-specific corpus artifacts.

\subsection{Semi-Supervised Self-Training}

In the semi-supervised stage, the evaluation assesses whether the teacher-student pseudo-labeling can further enhance the robustness. Two types of teacher models are used: a six-class L2-only teacher and an eight-class expanded teacher. Both teachers make predictions on an external audio pool, which excludes all samples used for supervised training, development, testing, and second-level external evaluation. Only the predictions with a confidence level of at least 0.95 are retained, and each prediction category is limited to a maximum of 500 samples.

\begin{table}[H]
\centering
\small
\setlength{\tabcolsep}{3pt}
\caption{Semi-supervised pseudo-label statistics.}
\label{tab:pseudo_label_stats}
\begin{tabular}{lccc}
\hline
Pseudo-label setting & Threshold & Max/class & Selected \\
\hline
Six-class teacher & 0.95 & 500 & 2606 \\
Eight-class teacher & 0.95 & 500 & 1037 \\
\hline
\end{tabular}
\end{table}

Table~\ref{tab:pseudo_label_stats} shows that the six-class teacher selects more pseudo-labeled samples than the eight-class teacher. However, the six-class teacher cannot represent British Isles English or Singaporean English, so external samples from these categories must be forced into one of the original six L2-ARCTIC labels. This introduces a clear risk of noisy pseudo-labels. The eight-class teacher is more compatible with the expanded label space, but the selected samples mainly concentrate in the two newly added external categories, thereby limiting the diversity of pseudo-label supervision.

Table~\ref{tab:semi_results} reports the final semi-supervised student results. As shown in Table~\ref{tab:semi_results}, in both six-class and eight-class settings, the student model does not outperform its supervised teacher. The six-class student decreases from 86.20\% to 83.00\% on the L2 test set and from 42.00\% to 38.00\% on the second-level external set. The eight-class student also decreases from 87.90\% to 85.00\% on the L2 test set and from 68.00\% to 58.00\% on the second-level external set. These results indicate that confidence-based pseudo-labeling alone is insufficient under domain mismatch.

\begin{table}[H]
\centering
\scriptsize
\setlength{\tabcolsep}{2pt}
\begin{tabular*}{\columnwidth}{@{\extracolsep{\fill}}lccc@{}}
\toprule
Model & \shortstack{L2 Acc.(\%)} & \shortstack{Ext. Acc.(\%)} & \shortstack{Level-2 Ext. (\%)} \\
\midrule
Sup. 6-cl Teacher & 86.20 & 71.05 & 42.00 \\
SSL 6-cl Student & 83.00 & 65.00 & 38.00 \\
Sup. 8-cl Teacher & 87.90 & 3.95 & 68.00 \\
SSL 8-cl Student & 85.00 & 4.00 & 58.00 \\
\bottomrule
\end{tabular*}
\caption{Semi-supervised student model results.}
\label{tab:semi_results}
\end{table}

Due to the differences in the label space and the composition of the external test samples, the external test accuracy rates of the six-class and eight-class settings cannot be directly compared. Therefore, the second-level external accuracy is taken as the main robustness indicator for the expanded setting.

\subsection{Discussion}

The experiments lead to four main findings. First, file-level evaluation substantially overestimates robustness. The replicated file-level baseline achieves 99.76\% internal L2 accuracy but only 44.52\% external accuracy, and the Round 1 file-level baseline reaches 100.00\% accuracy on L2-ARCTIC. In contrast, speaker-disjoint evaluation produces lower and more realistic estimates, suggesting that speaker overlap is a major source of performance inflation.

Second, external-only training does not transfer reliably back to L2-ARCTIC. Although external-only settings achieve high external comparable accuracy, they perform poorly on the official L2-ARCTIC evaluation. This suggests that external data alone may encode dataset-specific or source-specific patterns that do not align with the L2-ARCTIC accent categories.

Third, balanced external-data training is the most promising supervised configuration in this study, but it is not a complete solution. The expanded L2+external model improves over the L2-only baseline in all four Round 1 folds, and the Round 2 fold-0 expanded model improves second-level external accuracy from 42.00\% to 68.00\%. However, the L2 accuracy gains are modest, and the expanded Round 2 setting is not complete across all folds.

Fourth, confidence-based pseudo-labeling is unreliable under domain mismatch. The six-class teacher generates more pseudo-labeled samples but cannot represent external accent categories outside the original six labels. The eight-class teacher better matches the expanded label space, but its selected pseudo-labels are concentrated in British Isles English and Singaporean English. In both cases, the student model performs worse than the corresponding supervised teacher, indicating that high confidence does not necessarily imply correct or diverse supervision under cross-corpus mismatch.

Overall, the results support speaker-disjoint and cross-corpus evaluation as necessary components of robust accent-classification assessment. They also show that external data and pseudo-labeling must be handled carefully, because naive use of external samples can introduce source bias or noisy supervision.

\section{Conclusion and Future Work}

This study revisited DistilHuBERT-based accent classification and conducted experiments under three evaluation conditions: file-level, speaker-disjoint, and cross-corpus evaluation settings. The results showed that when the speakers in the training set and the test set overlap, the file-level evaluation would significantly overestimate the robustness; while the speaker-disjoint evaluation could more strictly reflect the generalization ability for speakers never seen before.

The experiments also show that external-only training does not reliably transfer back to L2-ARCTIC. In the evaluation configuration, balanced external data training is the most effective supervised setting, but its improvement effect is still limited. The semi-supervised pseudo-labeling method further demonstrated that in the case of domain mismatch, depending exclusively on  confidence-based filtering is insufficient.

Future work should improve the quality and coverage of external accent labels, evaluate stronger cross-domain adaptation methods, and examine more reliable semi-supervised strategies, including confidence calibration, consistency regularization, and teacher ensembling. Conducting a larger-scale evaluation of real user recordings will also more effectively test practical robustness. The confidence-thresholding approach is intended as a simple diagnostic semi-supervised baseline rather than a comprehensive semi-supervised learning framework.

\section*{Limitations}

This study has several limitations. Firstly, L2-ARCTIC only includes 24 speakers, with 4 speakers in each of the different first-language backgrounds. Although the speaker-disjoint protocol is more stringent than file-level splitting, the number of speakers in each accent category is still limited. Second, the external corpora vary in terms of recording conditions, label definitions, speaker demographics, and collection procedures. These differences make cross-corpora evaluation meaningful, but they inevitably lead to mismatches between the datasets.

Third, since not all the extended and semi-supervised settings were completed across all four speaker-disjoint folds, the results of Round 2 extension and pseudo-labeling should be regarded as diagnostic rather than statistically conclusive. Fourth, this study only evaluated the main backbone network based on DistilHuBERT.  Finally, in the pseudo-labeling stage, confidence filtering and per-class upper limits were used, but no confidence calibration, consistency regularization, or teacher ensembling were included. These methods may be necessary for achieving more reliable semi-supervised learning in the case of domain mismatch.

In addition, the present study is restricted to English accent classification. Whether the same evaluation effects generalize to larger multilingual or cross-lingual accent datasets remains to be investigated.



\bibliography{custom}

\end{document}